\documentclass[twocolumn]{jpsj2} 
\newcommand*{\sma}{Sm-\textit{A}}
\newcommand*{\IA}{\textit{I}-\textit{A}}
\newcommand*{\bmr}{\boldsymbol{r}}
\newcommand*{\bmu}{\hat{\boldsymbol{u}}}

\title{Smectic-\textit{A} Free Standing Film of Lennard-Jones
Spherocylinder Model}

\author{Masashi Torikai\thanks{E-mail address: torikai@phen.mie-u.ac.jp}}

\inst{Department of Physics Engineering, Faculty of Engineering, \\
Mie University, Tsu, 514-8507}

\abst{
 A spherocylinder-like molecule with a Lennard-Jones type interaction is 
 proposed as a model of smectic-\textit{A} (\sma) liquid crystals, which
 can form a free-standing film.
 By means of Gibbs ensemble simulations, the isotropic, nematic, and
 \sma{} phases of the model fluid are found to coexist with a vapor
 phase; 
 and the coexistence conditions of the liquid crystal phases with the
 vapor phase are determined.
 For a set of the interaction-parameters of the model molecule, the
 \sma{} free-standing film is produced below the bulk
 isotropic\textendash\sma{} phase transition temperature by using Monte
 Carlo simulations. 
 The film tension of the \sma{} free-standing film is calculated and its
 dependencies on the temperature and on the number of molecules are
 discussed. 
}

\kword{liquid crystal, smectic-\textit{A} phase, free standing film,
layer-thinning transition, spherocylinder, Lennard-Jones potential,
Gibbs ensemble simulation, Monte Carlo simulation}

\begin{document}
\maketitle

\section{\label{sec:introduction}Introduction}
Liquid crystals (LCs) in a smectic-\textit{A} phase (\sma) form a free
standing film (FSF) or freely suspended film.
The FSF is stable below the bulk phase transition temperature between
the \sma{} and lower ordered phase.
The FSF consists of multiple smectic layers and thus the film thickness
is approximately integral multiple of a single layer thickness.
Because the FSF can be very thin up to the two-layer thickness, it may
be strongly affected by its surfaces;
therefore the FSF is an appropriate system for us to investigate the
effects of free surfaces on an ordered phase.
A surface-enhancement of \sma{} order (SESO) observed in
experiments~\cite{Tweet1990,Holyst1991} is a phenomenon due to the
surface effect.
The SESO arises since the layer fluctuation of the \sma{} phase is
suppressed at the free surface by the surface
tension~\cite{Tweet1990,Holyst1991,Holyst1990}. 
A so-called layer-thinning (LT) transition~\cite{Stoebe1994,
Johnson1997, Pankratz1998, Pankratz1999, Pankratz2000,Veum2005,
Demikhov1995, Mol1998}, which occurs in fairly rare \sma{} materials, is 
another phenomenon resulting from the strong surface
effect. 
The \sma{} FSFs of LT materials can exist even at the higher temperature
than the bulk isotropic-liquid\textendash\sma{} (\IA) phase transition
temperature $T_{IA}$~\cite{Stoebe1994, Johnson1997,
Pankratz1998, Pankratz1999, Pankratz2000, Veum2005} or than the bulk
nematic\textendash\sma{} phase transition
temperature $T_{NA}$~\cite{Demikhov1995, Mol1998}.
The FSFs remain above $T_{IA}$ or $T_{NA}$ but the upper bound of the
number of layers is a decreasing function of the temperature, so that
the film thickness decreases layer-by-layer as the temperature increases.
Such layer-by-layer transitions result from a nucleation and growth of a
dislocation loop at the innermost layer; i.e., the melt of the innermost
\sma{} layer followed by the escape of LC molecules in the layer into a
meniscus at the film border~\cite{Pankratz1999,Pankratz2000,Mol1998}. 
The stability above $T_{IA}$ or $T_{NA}$ and
the melt of the innermost layer indicate that the \sma{} order in LT
materials is emphasized by the interfaces more strongly than in ordinary
\sma{} materials. 

Although the \sma{} FSFs are major substances in experiments, there are
no molecular models which exhibit the \sma{} FSF.
The well-known Gay-Berne (GB) model, which is an anisotropic molecule
with a spheroid-like intermolecular interaction, has isotropic
liquid($I$), nematic ($N$), \sma, and smectic-$B$
phases~\cite{Brown1998}; 
the GB model is frequently used as a model of uniaxial LC molecules.
By using sets of model-parameters with which the GB model fluid exhibits
vapor ($V$), $I$, and $N$ phases, it was shown that the GB fluid forms a
nematic FSF in equilibrium with the vapor 
phase~\cite{Martin1997,Mills1998}.
The \sma{} FSF, however, has not been found in GB fluids.
The GB model also has a disadvantage that it does not allow the
intuitive understanding of the relation between model-parameters and the
resulting potential since the GB potential depends on the parameters in
a complicated way. 
It was shown~\cite{Raton1996,Raton1997} by means of the density
functional theory (DFT) that the model fluid consisting of
hard-spheroids with attractive anisotropic Yukawa potential forms a
\sma{} FSF.
In a restricted region of potential parameters, the model \sma{} FSF
exhibits the SESO, as in the experiments on ordinary
\sma{}~\cite{Tweet1990,Holyst1991} and on the LT
materials~\cite{Pankratz2000}. 
Furthermore, refs.~\citen{Raton1996} and \citen{Raton1997} indicate that 
the model exhibits the LT transition in the proximity of a
$V$\textendash$I$\textendash$N$\textendash\sma{} quadruple point.
These DFT results are important to show the relation between the
intermolecular interaction and the stability of the FSF.
However, there always remains possibility that a foreseen phase is
stable, because only a limited number of phases is under consideration
in the DFT.

The main purpose of the present paper is to make a pair potential of a
LC molecule such that the LC forms the \sma{} FSF and to
investigate physical properties of the LC model fluid.
The new LC model is introduced in \S~\ref{sec:models}. 
By using Gibbs-ensemble simulations, in \S~\ref{subsec:bulk}, the model
fluid is shown to have $I$, $N$, and \sma{} phases in coexistence with
the vapor phase.  
In \S~\ref{subsec:film}, it is shown that the model fluid forms a
\sma{} FSF for a set of interaction-parameters.
It is found that the film is unstable above $T_{IA}$, and
hence the new LC model cannot be a model of LT materials.
However, the film tension of the LC model shows some resemblance to that 
of LT materials.
Such a resemblance is discussed in \S~\ref{sec:Conclusions}.
Some concluding remarks are also presented in \S~\ref{sec:Conclusions}. 

\section{\label{sec:models}Lennard-Jones Spherocylinder Model}
For the purpose to investigate the \sma{} FSF, it is necessary to find
first a molecular model which exhibits a \sma{} phase in a bulk system. 
Furthermore, in order to understand the relation between properties of
the pair potential and resulting phases, it is preferable that all the 
terms in the potential function of the model can be readily understood.

In this paper, I make a molecular interaction based on a hard
spherocylinder (SPC), with which the model fluid is known to exhibit the
\sma{} phase~\cite{Mcgrother1996}. 
The SPC is a cylinder capped with two hemispheres;
the diameter $D$ of the cylinder and hemispheres and the length of the
cylinder $L$ determine the shape of the SPC.
\begin{figure}
 \begin{center}
  \includegraphics[clip, width=7.5cm]{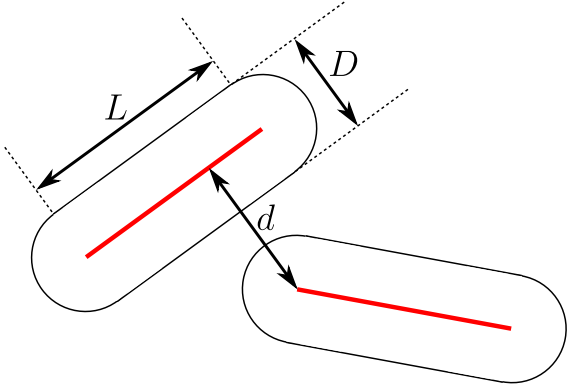}
 \end{center}
 \caption{\label{fig:spherocylinders}%
            (color-online)
            Two dimensional representation of spherocylinders. 
            The equidistant surface (the distance is $D/2$) from a line
            segment of length $L$ is the surface of a spherocylinder. 
            Therefore the interaction between two hard spherocylinders
            is a function of $d$, the distance between the line segments
            defining the spherocylinders.
            The new interaction between the soft spherocylinders
            introduced in this work is also defined as a function of $d$
            (see details in text). 
 }
\end{figure}
Because the surface of a SPC is the equidistant surface from the line
segment defining the center of the cylinder (see
Fig.~\ref{fig:spherocylinders}), the pair potential between two hard
SPCs is expressed using $d$, the distance between two line segments:
The pair potential of the hard SPCs is infinite in the case that
$d < D$ and otherwise zero.
The interaction between the new model molecules, named Lennard-Jones
spherocylinder (LJ-SPC), is also a function of $d$ and is a simple
extension of the hard SPC model to a soft molecule:
\begin{equation}
 V(\bmr_{12}, \bmu_{1}, \bmu_{2})
  =
  4V_{\text{o}}(\hat{\bmr}_{12}, \bmu_{1}, \bmu_{2})
  \left\{
   \left(
    \frac{D}{d}
   \right)^{12}
   -
   \left(
    \frac{D}{d}
   \right)^{6}
  \right\}, \label{eq:LJSPC}
\end{equation}
where $\bmr_{12} = \bmr_{2} - \bmr_{1}$;
$\bmr_{i}$ and $\bmu_{i}$ indicate, respectively, the center of
gravity and the unit vector along the symmetry axis of the $i$th
molecule;
$\hat{\bmr}_{12}$ is a unit vector $\bmr_{12}/|\bmr_{12}|$.
The function $V_{\text{o}}$ introduces an additional orientational
dependence:
\begin{multline}
 V_{\text{o}}(\hat{\bmr}_{12}, \bmu_{1}, \bmu_{2})
 =
 \varepsilon_{1}
 +
 \varepsilon_{2}
 P_{2}(\bmu_{1}\cdot\bmu_{2})\\
 +
 \frac{\varepsilon_{3}}{2}
 \left\{
 P_{2}(\hat{\bmr}_{12}\cdot\bmu_{1})
 +
 P_{2}(\hat{\bmr}_{12}\cdot\bmu_{2})
 \right\}, \label{eq:Vo}
\end{multline}
where $P_{2}(x)=(3x^{2} - 1)/2$ is the Legendre polynomial of second
order.
The terms in eq.\eqref{eq:Vo} correspond to the leading order terms in a
general spherical harmonic expansion of a function of $\hat{\bmr}_{12}$,
$\bmu_{1}$, and $\bmu_{2}$.
Similar expansion form of the anisotropic pair potential has been used
in refs.~\citen{Mederos1989, Mederos1992, Somoza1994, Somoza1995} with a
modified atomic LJ potential, and in refs.~\citen{Raton1996} and
\citen{Raton1997} with an anisotropic Yukawa attractive potential.
The first term in eq.\eqref{eq:Vo} gives uniform contribution and no
additional orientational dependence on the pair potential.
Thus, with $\varepsilon_{2} = \varepsilon_{3} = 0$, the molecule has
only the SPC-like orientational dependence resulting from the use of the
distance $d$.
In the case that $\varepsilon_{2} \geq 0$, the second term, which is a
Maire-Saupe-type interaction, makes parallel molecules stable and it
contributes to the nematic ordering. 
The third term makes, in the case that $\varepsilon_{3} \leq 0$, the
molecules prefer a side-by-side configuration rather than end-to-end
configuration, so that the term induces a \sma{} layering structure in
an orientationally ordered phase. 
\begin{figure}
 \begin{center}
  \includegraphics[clip, width=8.6cm]{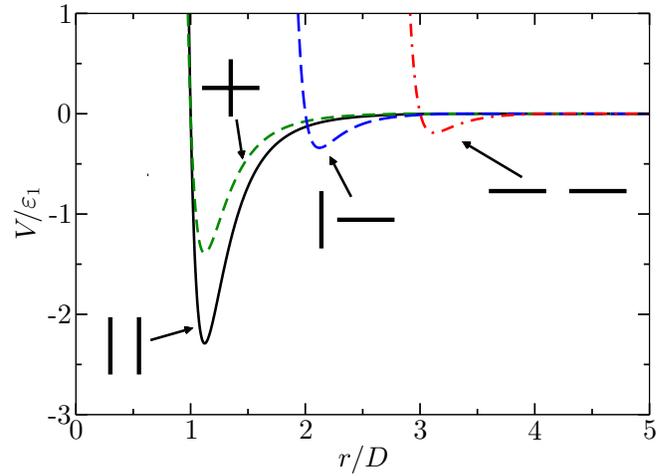}
 \end{center}
 \caption{\label{fig:potential}%
             (color-online)
            The Lennard-Jones spherocylinder interaction
            $V(\bmr_{12}, \bmu_{1}, \bmu_{2})$ for four typical
            configurations.
            The potential parameters are 
            $(\varepsilon_{1}, \varepsilon_{2}, \varepsilon_{3}) = (1, 0.6, -1.4)$.
            Four pairs of segments shown in the figure indicate the
            configurations of molecules. 
 }
\end{figure}
The shapes of $V(\bmr_{12}, \bmu_{1}, \bmu_{2})$ for typical
configurations are shown in Fig.~\ref{fig:potential}. 

\section{\label{sec:simulation}Simulation}
\subsection{\label{subsec:bulk}Bulk phases}
The main purpose of this section is the determination of the phases
coexisting with the vapor phase. 
Hereafter energies are measured in units of $\varepsilon_{1}$.
The length-to-breadth ratio $L/D$ is $2$ throughout the paper.
Although there are isotropic and solid phases but the \sma{} phase does
not exist in the case that the hard SPC molecules with
$L/D = 2$~\cite{Mcgrother1996}, we expect that the additional
orientational interaction $V_{\text{o}}$ may induce the \sma{} phase. 
The parameters $\varepsilon_{2}$ and $\varepsilon_{3}$ are restricted in
the region in which $V_{\text{o}}$ is positive for any configuration of
molecules. 

In performing the following simulations, a cutoff for intermolecular
interactions is introduced by means of
shift-and-truncation~\cite{Frenkel2002}. 
The shifted-and-truncated potential is obtained by shifting
the original potential by an amount $\delta$, followed by the truncation
if the shifted potential is attractive and positive.
The shift used in the present work is $\delta/\varepsilon_{1} = 0.01$.
The resulting cutoff distance is by definition anisotropic, e.g., the
cutoff distance in units of $D$ is 4.075 in the end-to-end configuration
and 3.120 in the side-by-side configuration for
$(\varepsilon_{2}, \varepsilon_{3})=(0.6, -1.4)$.
The maximum cutoff distance among the models that I have investigated is
5.097 for $(\varepsilon_{2}, \varepsilon_{3})=(1.2, 0.0)$.

The Gibbs ensemble Monte Carlo (GEMC) simulation~\cite{Frenkel2002} is
used in this section since the GEMC simulation is appropriate to find a
coexistence condition of two phases for the following reason.
Two simulation boxes are used in the GEMC simulation.
At a temperature sufficiently below the critical temperature at which
the coexistence line of two phases ends, the spontaneous phase
separation occurs in the GEMC simulation and the two simulation boxes
are filled with these coexisting phases.
Since the interface, whose effects on coexistence properties are
non-negligible, does not exist in GEMC simulation, quite large systems
are not necessary for phase equilibrium calculations.
Hence we can reduce the computer time utilizing the GEMC method.

The details of the simulations are as follows.
In all simulation runs in this section, the total number of molecules is
512; 
the initial dimension of each simulation box is defined so that the
reduced density of molecules is 0.1.
Here the reduced density $\eta$ is defined as
$\eta=N/(\rho_{c}V)$ with the number of molecules $N$, the volume of the
simulation box $V$, and the hexagonal-close-packing density
$\rho_{c}D^{3}=2/(\sqrt{2} + \sqrt{3}L/D)$ of perfectly aligned hard
SPCs. 
Each simulation box is maintained to be a cube throughout the
simulation, and the periodic boundary condition is used.
In order to make a uniform and isotropic initial configuration, I
performed a simulation for several thousands of MC steps at
$\tilde{T} = k_{\text{B}}T/\varepsilon_{1} = 2.0$ ($k_{\text{B}}$ is the
Boltzmann constant, $T$ is the temperature, and $\tilde{T}$ is the
dimensionless temperature). 
The simulations were performed from higher temperature to lower
temperature;
the last configuration of a simulation at higher temperature was used
as the initial configuration of a lower temperature simulation.
In order to determine the upper and lower bounds of the transition
temperature, heating processes are also simulated around the transition
temperature, i.e., the last configuration at the lower temperature is
used as an initial configuration of a higher temperature simulation. 
The sets of potential parameters investigated in this work are 
$\varepsilon_{2} = \{0, 0.6, 1.2\}$ and
$\varepsilon_{3} = \{0, -0.5, -0.9, -1.2, -1.4\}$;
$\varepsilon_{3} = -1.2$ and $-1.4$ are omitted at $\varepsilon_{2} = 0$
and $1.2$ because $V_{\text{o}}(\hat{\bmr}_{12}, \bmu_{1}, \bmu_{2})$
takes a negative value depending on the configuration of molecules.

The GEMC simulation results of the phases which coexist with
vapor phase are summarized in the Table~\ref{table:phases}.
\begin{table}
 \caption{\label{table:phases}The phases coexisting with a vapor phase
 in the Gibbs ensemble method.
 The letters $I$, $N$, and $A$ denote the isotropic liquid, nematic,
 and \sma{} phases, respectively.}
 \begin{tabular}{cccccc}
  \hline
  \hline
  & $\varepsilon_{3} =  0$       & $-0.5$     &  $-0.9$ & $-1.2$ & $-1.4$\\
  \hline
  $\varepsilon_{2} =  0$ & $I$ & $I$ & $I$ & - & - \\
  $0.6$ & $N$, $A$ & $N$, $A$ & $I$, $A$ & $I$, $A$ & $I$, $A$ \\
  $1.2$ & $N$, $A$ & $A$ & $A$ & - & - \\
  \hline
  \hline
 \end{tabular}
\end{table}
These results show that $\varepsilon_{2}$ is necessary
for the orientational ordering.
The LJ-SPC model always exhibits the \sma{} phase if 
$\varepsilon_{2}$ is non-zero.
The third parameter $\varepsilon_{3}$ is responsible
whether the $I$ or $N$ phase is dominant.

\begin{figure}
 \begin{center}
  \includegraphics[clip, width=8.6cm]{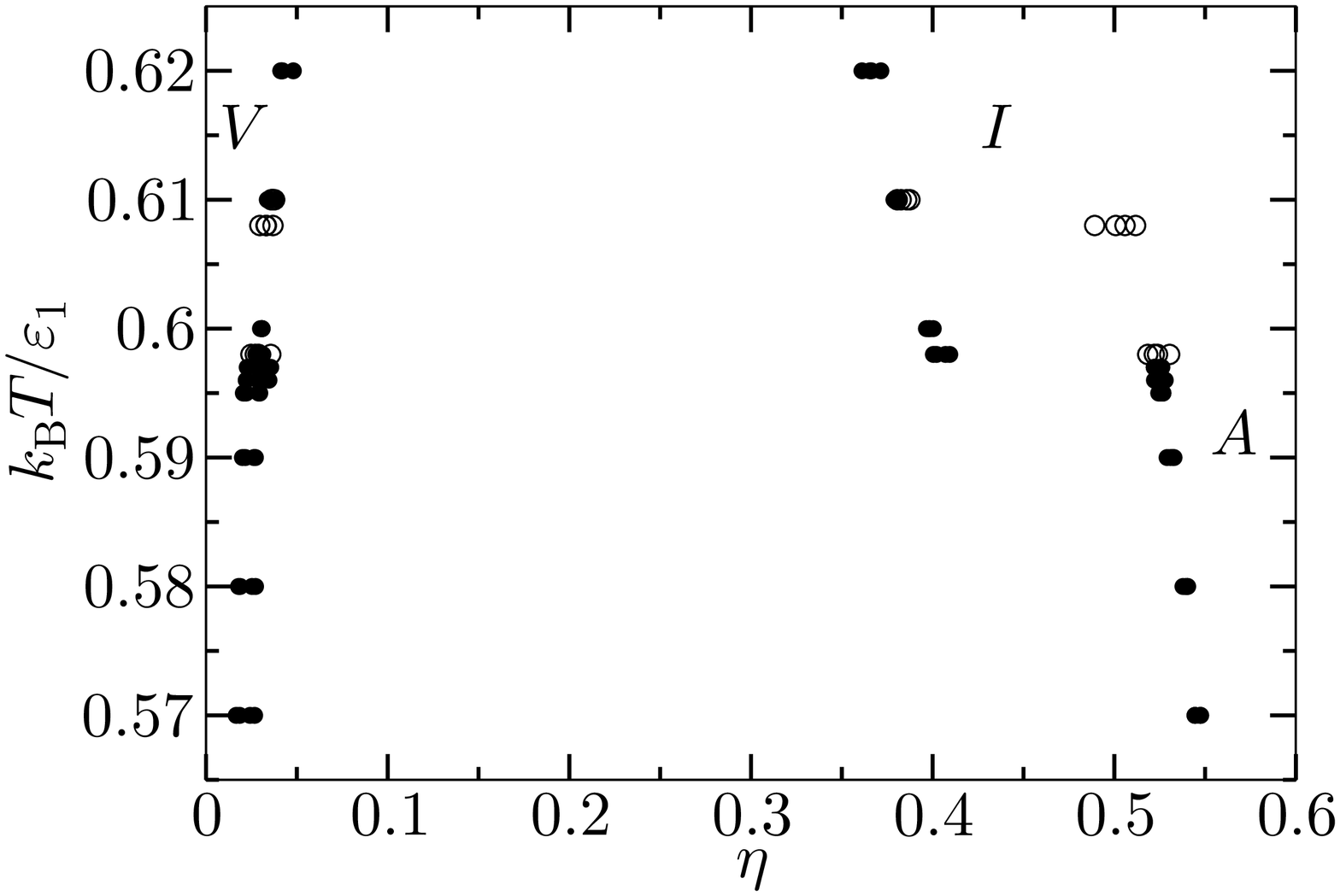}
  \includegraphics[clip, width=8.6cm]{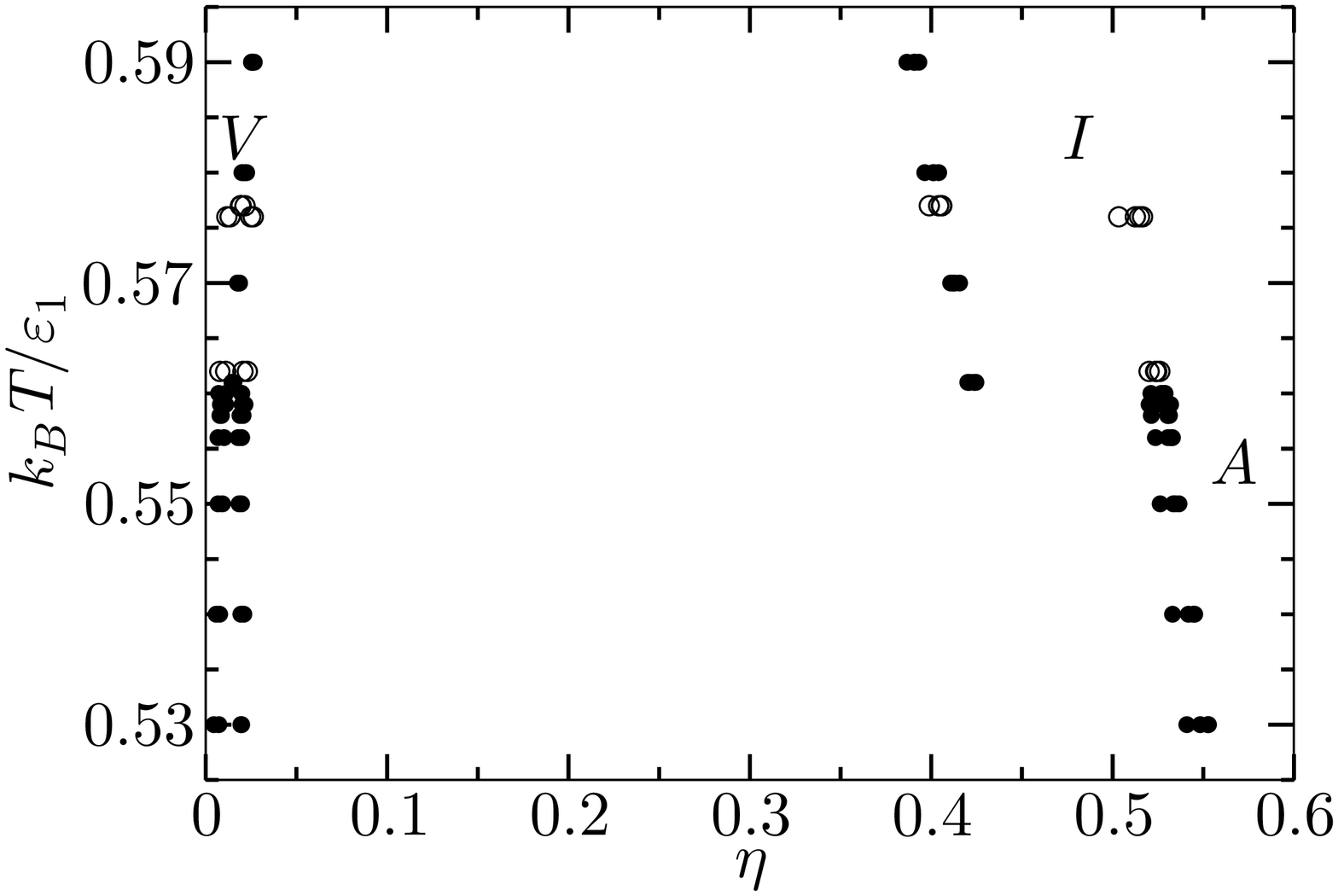}
 \end{center} 
 \caption{\label{fig:phaseDiagrams}%
             Phase diagram of LJ-SPC model with
            $(\varepsilon_{2}, \varepsilon_{3})=(0.6, -1.2)$ (upper
            panel) and $(0.6, -1.4)$ (lower panel).
            Solid and open circles denote the reduced density observed
            in the cooling process and heating process, respectively.
            The vapor, isotropic liquid and \sma{} phases are denoted by
            $V$, $I$  and $A$, respectively.
            Data points of reduced densities $\eta$ obtained from four
            independent simulation runs are plotted at each temperature.
 }
\end{figure}

The LJ-SPC model with parameters
$(\varepsilon_{2}, \varepsilon_{3}) = (0.6, -1.2)$ and $(0.6, -1.4)$,
with which the model fluid undergoes a direct
\IA{} phase transition, is of interest since the
most of LT transitions observed in such LCs~\cite{Stoebe1994,
Johnson1997, Pankratz1998, 
Pankratz1999, Pankratz2000, Veum2005}. 
The phase diagrams of the LJ-SPC model with these parameters are shown
in Fig.~\ref{fig:phaseDiagrams}. 
Each data point plotted in Fig.~\ref{fig:phaseDiagrams} is an average
of $\eta$ over $5\times10^{5}$ MC steps. 
The transition temperature $\tilde{T}_{IA}$ of the LJ-SPC model is,
according to the Fig.~\ref{fig:phaseDiagrams},
$0.597 < \tilde{T}_{IA} < 0.610$ for
$(\varepsilon_{2}, \varepsilon_{3}) = (0.6, -1.2)$ and
$0.560 < \tilde{T}_{IA} < 0.577 $ for
$(\varepsilon_{2}, \varepsilon_{3}) = (0.6, -1.4)$.
For both sets of parameters, the \sma{} layer spacing at $T_{IA}$ is
approximately $3$ in units of $D$, i.e., approximately the molecular
length $L + D$.

\subsection{\label{subsec:film}Smectic-\textit{A} free standing film}
In this section, I determine a set of parameters for the existence of
the \sma{} FSFs of LJ-SPC fluid, and investigate their properties.
I also establish whether the LJ-SPC fluid corresponds to the LT
materials.
However, since the nucleation and growth of the dislocation loop are
suppressed in a finite system~\cite{Holyst1994,Holyst1995}, we cannot
expect to observe the LT without enormous amount of time.
Thus I do not intend to observe the nucleation and growth of the
dislocation loop, but focus only on the static properties characteristic
to the LT materials. 

For simulations of the LC film in the $NVT$ ensemble, I used an ordinary
Metropolis method. 
The simulation box is a three-dimensional box with dimensions
$(L_{x}, L_{y}, L_{z}) = (15, 15, 50)$ in units of $D$ with periodic
boundary conditions in all directions.
For the simulations in the present section, I used two kinds of initial
state, which are prepared in the following ways:
\textit{\sma{} initial state}:
slicing the \sma{} and vapor phase obtained from the GEMC simulation in
\S\ref{subsec:bulk} and sandwiching the \sma{} slice between the two
vapor slices;  
\textit{uniform initial state}:
equilibrating the fluid at sufficiently high temperature
($\tilde{T} = 2.0$) for several thousands of MC steps until the fluid
becomes uniform and isotropic. 

In the case that $(\varepsilon_{2}, \varepsilon_{3}) = (0.6, -1.2)$ at
the lower bound of the transition temperature $\tilde{T} = 0.597$ a
simulation from the \sma{} initial state shows that the initial \sma{}
FSF melts into an isotropic film after $5\times10^{5}$ MC steps.
From the uniform initial state at the same temperature, although the
uniform isotropic fluid condenses into one or two films in equilibrium
with the vapor phase, the condensed phase is still isotropic.
The similar results are observed at
$\tilde{T} = 0.580 < \tilde{T}_{IA}$, hence \sma{} film is not stable
in $\tilde{T} \geq 0.580$.
At $\tilde{T} = 0.570$, the \sma{} FSF is stable.
These results mean that the melting temperature of the \sma{} FSF is
fairly lower than the bulk \IA{} phase transition temperature for
$(\varepsilon_{2}, \varepsilon_{3}) = (0.6, -1.2)$.

\begin{figure}
 \begin{center}
  \includegraphics[clip, width = 4.cm]{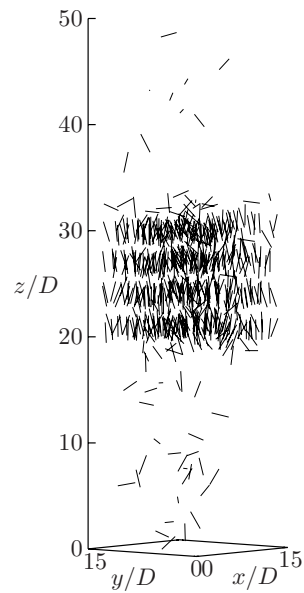}
 \end{center} 
 \caption{\label{fig:4layer}%
             A snapshot of a 4-layer \sma{} FSF at
            $\tilde{T}=\tilde{T}_{IA}=0.560$ with the potential
            parameters
            $(\varepsilon_{2}, \varepsilon_{3}) = (0.6, -1.4)$.
            Each molecule is drawn as a segment of length $2D$.
 }
\end{figure}

In the case $(\varepsilon_{2}, \varepsilon_{3}) = (0.6, -1.4)$, from the
\sma{} initial state, the \sma{} film does not melt at the lower bound
of the phase transition temperature $\tilde{T} = 0.560$, unlike in the
case $\varepsilon_{3} = -1.2$.
Furthermore, in the simulations at $\tilde{T} = 0.560$ from a uniform
initial state, the LJ-SPC fluid spontaneously condenses into one or two
\sma{} FSFs. 
One of a typical snapshot of the \sma{} FSF is shown in
Fig.~\ref{fig:4layer}. 
The \sma{} FSFs with 3, 4, 5, and 6 layers are stable, but 1- and
2-layer \sma{} FSFs show strong fluctuation and seem to be unstable.
The \sma{} layer spacing is almost the same as that of the bulk \sma{}
phase. 
The \sma{} FSF prepared at $\tilde{T} = 0.560$ melts into isotropic film
if it is heated up to $\tilde{T} = 0.567$.
Since the hysteretic region of the \sma{} FSF is included in that of the
bulk \sma{}, we can expect that the \IA{} phase transition temperature
of the \sma{} FSF is lower than or in the vicinity of the bulk
$\tilde{T}_{IA}$. 
Hence we can conclude that the LJ-SPC model with
$(\varepsilon_{2}, \varepsilon_{3}) = (0.6, -1.4)$ is similar to the
normal \sma{} materials, which can produce the FSF in the vicinity of
$\tilde{T}_{IA}$, but the LJ-SPC model does not correspond to the LT
materials, which exhibit the \sma{} FSF at higher temperature than the
bulk $\tilde{T}_{IA}$. 
In the rest of the present paper, I investigate the properties
of the \sma{} FSF of the LJ-SPC model with 
$(\varepsilon_{2}, \varepsilon_{3}) = (0.6, -1.4)$.

In the case that the \sma{} initial state is used, the resulting condensed
phase can be one or two films depending on the initial configuration and
random-number sequence used in MC simulations.
When the double-film state was obtained, I repeatedly performed the
simulation run using different random-number sequences until a
single-film state was obtained.
The following numerical results are obtained from single-film states
and the data from double-film states are omitted.

\begin{figure}
 \begin{center}
  \includegraphics[clip, width = 8.6cm]{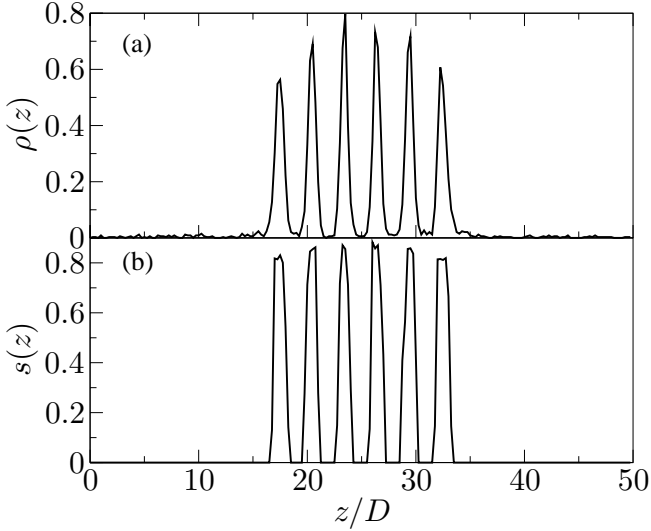}
 \end{center}
 \caption{\label{fig:localOrder}%
 (a)Local density $\rho(z)$ and (b) local orientational order $s(z)$.
 The set of interaction parameters is
 $(\varepsilon_{2}, \varepsilon_{3}) = (0.6, -1.4)$ and the temperature
 is $\tilde{T} = 0.556$.
 }
\end{figure}

In order to check the SESO, I calculated a local density and local
orientational order parameter. 
Here the local density $\rho(z)$ is the number of molecules
in $[z, z + \Delta z]$ divided with a volume $L_{x}L_{y}\Delta z$.
The local orientational parameter $s(z)$ is defined as the maximum
eigenvalue of an order parameter tensor $Q(z)$ with components
\begin{equation}
 Q_{\mu \nu}(z)
  =
  \frac{1}{N_{z}}
  \sum_{i}^{(z)}
  \langle \hat{u}_{i \mu} \hat{u}_{i \nu} \rangle
  - \frac{1}{3}\delta_{\mu \nu},
  \quad  \mu, \nu = x, y, z
\end{equation}
where the summation is performed over all the molecules within
$[z, z + \Delta z]$ and $N_{z}$ denotes the number of molecules within
$[z, z + \Delta z]$.
In this paper, $\Delta z/D$ is $0.25$.
Figure~\ref{fig:localOrder} shows the local density and local
orientational order parameter along the $z$ axis at $\tilde{T} = 0.556$.
Both the local density and local orientational order have maximum peaks
at innermost layers and gradually decrease to the free-surfaces, i.e., 
the free-surfaces reduce the \sma{} order in the LJ-SPC model.
This is in contrast to the fact that the SESO is observed in
experiments~\cite{Pankratz2000,Tweet1990,Holyst1991}.

\begin{figure}
 \begin{center}
  \includegraphics[clip, width = 8.6cm]{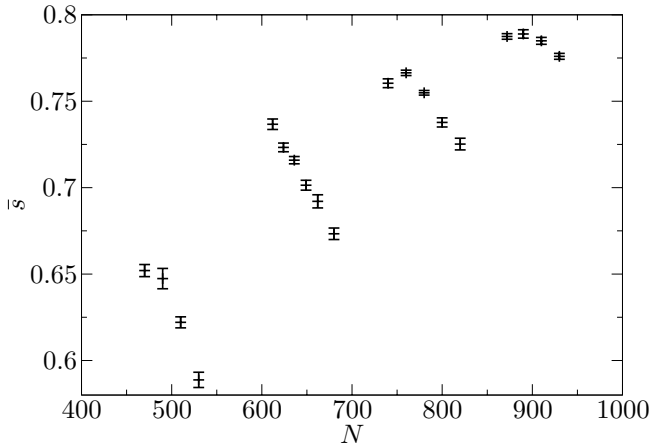}
 \end{center}
 \caption{\label{fig:P2vsN}%
            The number of molecules ($N$) dependence of the averaged
            orientational order parameter $\bar{s}$ at
            $\tilde{T}=0.556$. 
            The observed \sma{} FSFs consist of 3, 4, 5, and 6 layers in
            the case that $N \in [470, 530]$, $[612, 680]$,
            $[740, 820]$, and $[872, 930]$, respectively.  
 }
\end{figure}

The disordering caused by the existence of free surfaces can also be
seen in the behavior of the averaged orientational order parameter
$\bar{s}:=\int\mathrm{d}z \, s(z)/L_{z}$.
The $N$ dependence of $\bar{s}$ is shown in Fig.~\ref{fig:P2vsN}.
The data in this figure are obtained using ten thermal averages of
$\bar{s}$ calculated from ten runs of $2\times10^{6}$ MC steps for each
$N$, following an equilibration of $10^{7}$ MC steps.
The standard error of the mean is obtained from these ten averages
with an assumption that the error distribution is a Gaussian.
Figure~\ref{fig:P2vsN} shows that $\bar{s}$ increases as the number of
layers increases.
This fact is another evidence for that the disordering effect of free
surfaces strongly affects the thinner \sma{} FSFs and makes their order
parameter small.

The rest of this section is dedicated to the calculation of the film
tension of the \sma{} FSF.
The film tension:
\begin{equation}
 \gamma
  =
  \int_{0}^{L_{z}} \mathrm{d} z
  \left[p_{\text{n}}(z) - p_{\text{t}}(z)\right], \label{eq:filmTension}
\end{equation}
is derived using the normal and tangential components of the pressure
tensor~\cite{Martin1997,Mills1998} defined respectively as:
\begin{align}
 p_{\text{n}}(z) &=
 \rho(z) \tilde{T} -
 \frac{1}{2V_{\text{c}}} \sum_{i, j}{}' z_{ij}
 \frac{\partial V_{ij}}{\partial z_{ij}} \\
 p_{\text{t}}(z) &=
 \rho(z) \tilde{T} -
 \frac{1}{4V_{\text{c}}} \sum_{i, j}{}'
 \left(
 x_{ij}
 \frac{\partial V_{ij}}{\partial x_{ij}}
 +
 y_{ij}
 \frac{\partial V_{ij}}{\partial y_{ij}}
 \right), \label{eq:pnANDpt}
\end{align}
where $(x_{ij}, y_{ij}, z_{ij}) = \bmr_{j} - \bmr_{i}$, $V_{ij}$
is the potential energy between the $i$th and $j$th molecules, and
$V_{\text{c}}$ is the volume of a slice $[z, z + \Delta z]$, i.e.
$V_{\text{c}} = L_{x}L_{y}\Delta z$.
The summation in eq.~\eqref{eq:pnANDpt} is done if at least one of the
$i$th and $j$th molecules is in the slice.

\begin{figure}
 \begin{center}
  \includegraphics[clip, width = 8.6cm]{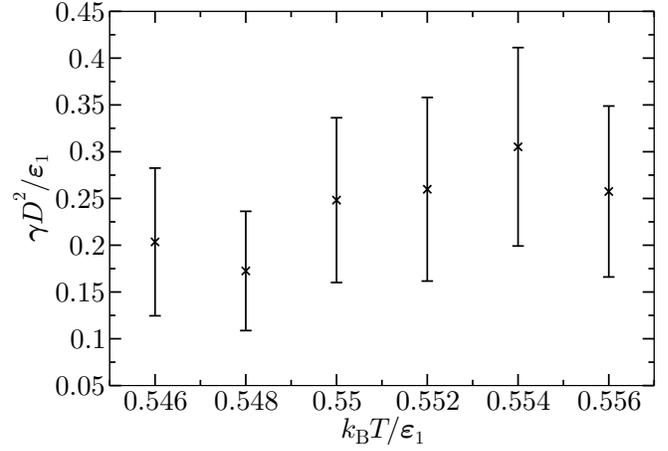}
 \end{center} 
 \caption{\label{fig:filmTensionT}%
             The film tension $\gamma$ of \sma{} FSF with
            $(\varepsilon_{2}, \varepsilon_{2}) = (0.6, -1.4)$.
            The number of molecules is 612, with which the \sma{} FSF
            consists of 4 layers.
 }
\end{figure}

I measured first the temperature dependence of the film tension at some
temperatures below the $T_{IA}$.
The data used in this calculation are from MC runs performed in the same
manner to the $\bar{s}$ calculation explained above (i.e., $10^{7}$ MC
steps for equilibration and ten runs of $2\times10^{6}$ MC steps for
obtaining thermal averages).
The temperature dependence of the average film tension and its standard
error of the mean are shown in Fig.~\ref{fig:filmTensionT};
the standard error of the mean is rather large as a result of
the strong fluctuation of the film tension in simulation runs.
Although it is difficult to state a definite conclusion because of the
large standard error, the dependence of the film tension on $T$ is very
weak.

\begin{figure}
 \begin{center}
  \includegraphics[clip, width = 8.6cm]{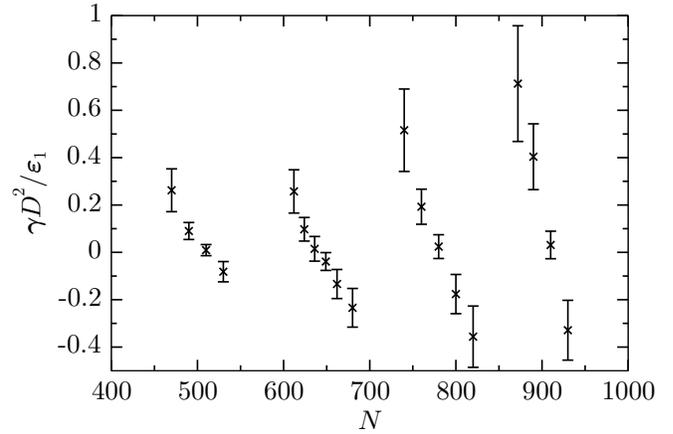}
 \end{center} 
 \caption{\label{fig:filmTensionN}%
             The film tension $\gamma$ of \sma{} FSF with
            $(\varepsilon_{2}, \varepsilon_{2}) = (0.6, -1.4)$ at
            $\tilde{T} = 0.556$. 
            The simulation runs used in this figure are the same as
            those in Fig.~\ref{fig:P2vsN}. 
 }
\end{figure}

From the calculation of the film tension with many initial
configurations, it is found that the film tension strongly depends on
the number of molecules $N$.
The $N$ dependence of the film tension is shown
in Fig.~\ref{fig:filmTensionN}.
The set of simulation runs used in this figure is the same as that used
in calculating $\bar{s}$ in Fig.~\ref{fig:P2vsN}.
The Fig.~\ref{fig:filmTensionN} clearly shows that $\gamma$ decreases as
$N$ increases if the number of layers remains constant, and $\gamma$
increases discontinuously at the point at which the number of layers
increases. 

\section{\label{sec:Conclusions}Concluding Remarks}
In this paper, I proposed a new LC model (LJ-SPC model) and showed that
the model exhibits the bulk \sma{} phase and \sma{} FSF.
The main features of the LJ-SPC model is the SPC-symmetric part and
additional orientational part $V_{\text{o}}$.
The former is introduced to increase the stability of the \sma{} phase.
The latter has the Maire-Saupe type interaction, which is often used in
the molecular theoretical studies~\cite{McMillan1971, Kobayashi1970,
Kobayashi1971} to induce the orientational order, and the term that 
governs the coupling between translational and orientational degrees of
freedom.
In contrast to the models \cite{McMillan1971, Kobayashi1970,
Kobayashi1971} neglecting the anisotropic repulsive part and the
coupling between translational and orientational degrees of freedom, the
LJ-SPC model exhibits the \sma{} phase without resorting to the mean
field approximation.

From GEMC simulations in the bulk, the parameter
$\varepsilon_{2}$ in eq.~\eqref{eq:Vo} is found to make a dominant
contribution to the orientational order. 
The third parameter $\varepsilon_{3}$, which is set negative in this
paper, contributes to diminish the orientational order in the bulk so
that the $N$ phase disappears if $|\varepsilon_{3}|$ is large.
The $\varepsilon_{3}$, on the other hand, plays a crucial role to
stabilize the \sma{} FSF as shown in \S~\ref{subsec:film}.
This is because the negative $\varepsilon_{3}$ term makes the
side-by-side configuration stable and the end-to-end configuration
relatively unstable, so that the perpendicular alignment becomes to be
preferred at a free surface;
hence the \sma{} film becomes stable.

It is clear that the LJ-SPC molecule is not appropriate as a model of
the LT materials since the \sma{} FSF of the LJ-SPC is unstable above
$T_{IA}$.
Furthermore, the experiments~\cite{Pankratz2000} show that the SESO is
essential for the LT, but the free-surface of the LJ-SPC fluid reduces
the \sma{} ordering.
Hence we can conclude that the LJ-SPC model does not satisfy the
condition to be a model of LT materials.

The LJ-SPC model, however, bears a resemblance to the LT materials in
the behavior of the film tension, which was precisely measured in 
ref.~\citen{Veum2005}.
The weak $T$-dependence of $\gamma$ below the $T_{IA}$ observed in
ref.~\citen{Veum2005} is also observed in Fig.~\ref{fig:filmTensionT} in
the present paper.
Above the $T_{IA}$ the film tension observed in ref.~\citen{Veum2005}
increases linearly with the temperature difference $T - T_{IA}$ at the
constant number of \sma{}\textendash layers, and discontinuously
decreases if the layer-by-layer thinning occurs.
Hence the \sma{} FSF of LT materials shows a sawlike dependence of the
film tension on the temperature.
The experiments in refs.~\citen{Mol1998} and \citen{Pankratz2000} showed
that the melt of the innermost layer of the \sma{} FSF followed by the 
escape of the molecules into the meniscus occurs at a single
layer-by-layer thinning transition. 
It is natural to assume that the number of molecules in the FSF
decreases monotonically as the temperature increases even in a
temperature region between two successive layer-by-layer transitions. 
If that is the case, since the tension of the LJ-SPC FSF decreases as
the number of particles decreases as shown in
Fig.~\ref{fig:filmTensionN}, we can assume that the sawlike
dependence on the temperature of the tension of \sma{} FSF in
ref.~\citen{Veum2005} is represented by the LJ-SPC FSF.

The length-to-breadth ratio $L/D$, which is fixed at $2$ in this paper,
affects the stability of the \sma{} phase.
The models with larger length-to-breadth ratio are important in
considering the realistic \sma{} LCs, since the real \sma{} LCs have
much larger length-to breadth ratio (e.g. the
4-cyano-4'-$n$-alkylbiphenyl ($n$CB) series exhibits \sma{} phase only
for $n \geq 8$).
In general, the larger the length-to-breadth ratio is, the more
stable the \sma{} phase becomes.
Hence we can expect that the bulk \sma{} phase and \sma{} FSF will be
stable for the LJ-SPC model not only with $L/D = 2$ but also with
$L/D > 2$.
In order to simulate the realistic \sma{}, we have to adopt larger
length-to-breadth ratios.
Then the required values of $\varepsilon_{2}$ and $|\varepsilon_{3}|$
for the existence of the bulk \sma{} and \sma{} FSF will be smaller than
those of the LJ-SPC fluid with $L/D = 2$.
That is, in the case of realistic \sma{} LCs, the higher order terms in
the spherical harmonic expansion $V_{\text{o}}$ will make a small
contribution to the stability of the \sma{} phase.

I add some comments about the absence of the SESO in the LJ-SPC fluid.
In the presence of a \textit{V}-\textit{A} interface, the surface
tension suppresses the layer
fluctuation~\cite{Tweet1990,Holyst1991,Holyst1990}, which results from
the Landau-Pierls instability. 
The results in \S IIIB, however, show that the effect of the surface
tension is not enough to promote the SESO.
The absence of the SESO in the present finite system may be attributed
to the absence of a long-wavelength fluctuation, which gives dominant
contribution to the total
layer-fluctuation~\cite{Holyst1991,Holyst1990}. 
Since the SESO is induced by the suppression of the fluctuation at the
surfaces, the underestimate of the fluctuation may result in the
omission of the SESO.
There is a possibility that the SESO arises in the LJ-SPC FSF by
including the long-wavelength fluctuation. 
The estimation of the effect of the long-wave length fluctuation on the
SESO is left for future studies.
In addition to such a hydrodynamic origin of the SESO, we can expect a
molecular origin of the SESO.
In order to promote the SESO, the authors in
ref.~\citen{Raton1996} and \citen{Raton1997} used a general spherical
harmonic expansion of the attractive anisotropic interaction, and
configured so that the terms of the interaction have the different ranges.
Since the contribution of the terms favoring the \sma{} configuration to
the free energy is proportional to the square of density gradient, these
terms effectively deduce the total free energy if their ranges are
sufficiently large to cover the interfacial region.
Indeed, in ref.~\citen{Raton1996} and \citen{Raton1997}, setting the
range of the terms favoring the \sma{} configuration largest, the authors 
showed by means of DFT that the SESO state has the minimum free energy.
By the same reason, there is a possibility that the LJ-SPC fluid 
exhibits the SESO by introducing the concept of different interaction
ranges.
This requires further investigations.

\begin{figure}
 \begin{center}
  \includegraphics[clip, width = 8.6cm]{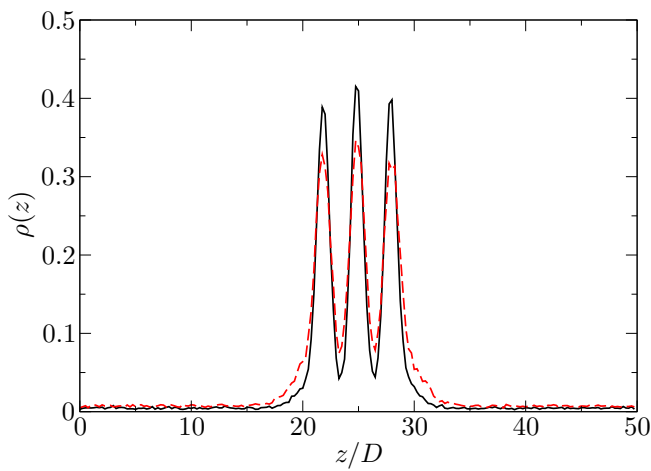}
 \end{center} 
 \caption{\label{fig:distortion}%
            (color online)
            The density profiles of 4-layer film with the number of
            molecules $N=470$ (black solid curve) and $530$ (red dashed
            curve). 
 }
\end{figure}
The negative film tension (i.e. interfacial free energy per unit area)
observed in Fig.~\ref{fig:filmTensionN} requires further consideration.
Since a negative interfacial free energy induces an expansion of the
interface, the area of the interface tends to diverge.
Therefore the negative film tension is inhibited to the thermodynamical
stable state.
However, in the present case, the area of the vapor and \sma{} interface
cannot diverge without a distortion.
Since the total free energy increases as a result of the distortion, the 
distortion will be stopped at a state at which the free energy of the
distortion and interfacial free energy balance each other.
The distortion of the film can be observed in Fig.~\ref{fig:distortion}
which shows the 3-layer density profiles of $N=470$ and $530$;
the peaks in the density profile of $N=530$ film is broader than that of
$N=470$ film due to the film distortion. 
The averaged orientational order parameter $\bar{s}$ is also useful to
estimate the distortion.
For a fixed number of layers, $\bar{s}$ is an essentially decreasing
function of the number of molecules as shown in Fig.~\ref{fig:P2vsN}.
This fact indicates that the \sma{} FSF is distorted as a result of the
negative interfacial free energy and therefore the orientational order
is reduced. 
Under conditions of the real experiment, however, the number of
molecules consisting the FSF may decrease until the film tension becomes 
positive since the molecules can flow into the meniscus.
Hence the distortion of the FSF may not be observed.

\section*{Acknowledgment}
The author acknowledges helpful discussions with Professor Mamoru
Yamashita.


\end{document}